\documentclass{optica-article}

\journal{opticajournal} 

\articletype{Research Article}

\usepackage{lineno, soul, color}

\begin{document}

\title{Contrary to widespread belief, the Fresnel zone plate outperforms the metalens at high NA.}

\author{Apratim Majumder,\authormark{1,$\dagger$} John A. Doughty,\authormark{1,$\dagger$} Tina H. Hayward,\authormark{1} Henry I. Smith,\authormark{2,3} and Rajesh Menon\authormark{1,4,*}}

\address{\authormark{1}Dept. of Electrical \& Computer Engineering, University of Utah, Salt Lake City UT 84112, USA\\
\authormark{2}Dept. of Electrical Engineering \& Computer Science, Massachusetts Institute of Technology, Cambridge MA 02139, USA.\\
\authormark{3}LumArray, Inc. Somerville MA 02143, USA.\\
\authormark{4}Oblate Optics, Inc. San Diego CA 92130, USA.\\
\authormark{$\dagger$}Equal contribution.}

\email{\authormark{*}rmenon@eng.utah.edu} 


\begin{abstract*} 
Rigorous simulations challenge recent claims that metalenses outperform conventional diffractive lenses, such as Fresnel Zone Plates (FZPs), in focusing efficiency at high numerical apertures (NAs). Across various lens diameters, FZPs exhibit a pronounced asymmetry in the shadow effect, leading to significantly higher focusing efficiency when optimally oriented. Extending this analysis, we show that conventional blazed gratings also surpass meta-gratings in efficiency. Since any linear optical element can be decomposed into local gratings, these findings broadly underscore the superiority of blazed structures over binary metastructures. Experimental characterization of an FZP with diameter = 3 mm, focal length = 0.2 mm operating at $\lambda$ = 634 nm confirms the dependence of efficiency on illumination direction. Our results emphasize the need for rigorous, direct comparisons between meta-optics and traditional diffractive optics to ensure accurate performance assessments.
\end{abstract*}

\section{Introduction}
The quintessential diffractive lens, discovered by Augustin Fresnel in 1866, acts like a hologram formed by the interference between a plane wave and a converging spherical wave.\cite{fresnel1996calcul} When implemented as a continuous phase structure, known as a Fresnel Zone Plate (FZP, sometimes referred to as the blazed Fresnel phase plate or the Fresnel lens), it efficiently converts an incident plane wave into a converging spherical wave with minimal thickness $=\lambda$/(n-1), where $\lambda$ is the wavelength and n is the refractive index.  At high numerical apertures (NAs), the focusing efficiency of diffractive lenses, such as FZPs, is known to degrade.\cite{prather2001performance,chao2005immersion} To mitigate this, optimization-based methods,\cite{menon2015nanophotonic, finlan1995efficient} now commonly referred to as \emph{inverse design}, have been applied since at least the early 1990s. 

While definitions vary, \emph{metalenses} are widely regarded as flat lenses with subwavelength features that are distinct from FZPs.\cite{menon2023inconsistencies, banerji2019imaging,mait2001selected} Over the past decade, metalenses have been praised for supposedly achieving \emph{higher focusing efficiency} than traditional diffractive lenses. \cite{arbabi2023advances, lalanne2017metalenses} Yet, to our knowledge, no rigorous validation supports these claims. Indeed, two recent studies directly comparing metalenses and FZPs with identical specifications showed the opposite: FZPs consistently outperformed metalenses,\cite{wang2024grayscale, vanmol2024fabrication} although these comparisons were limited to NA $\leq$ 0.6.

\begin{figure}[htbp]
\centering\includegraphics[width=11.4cm]{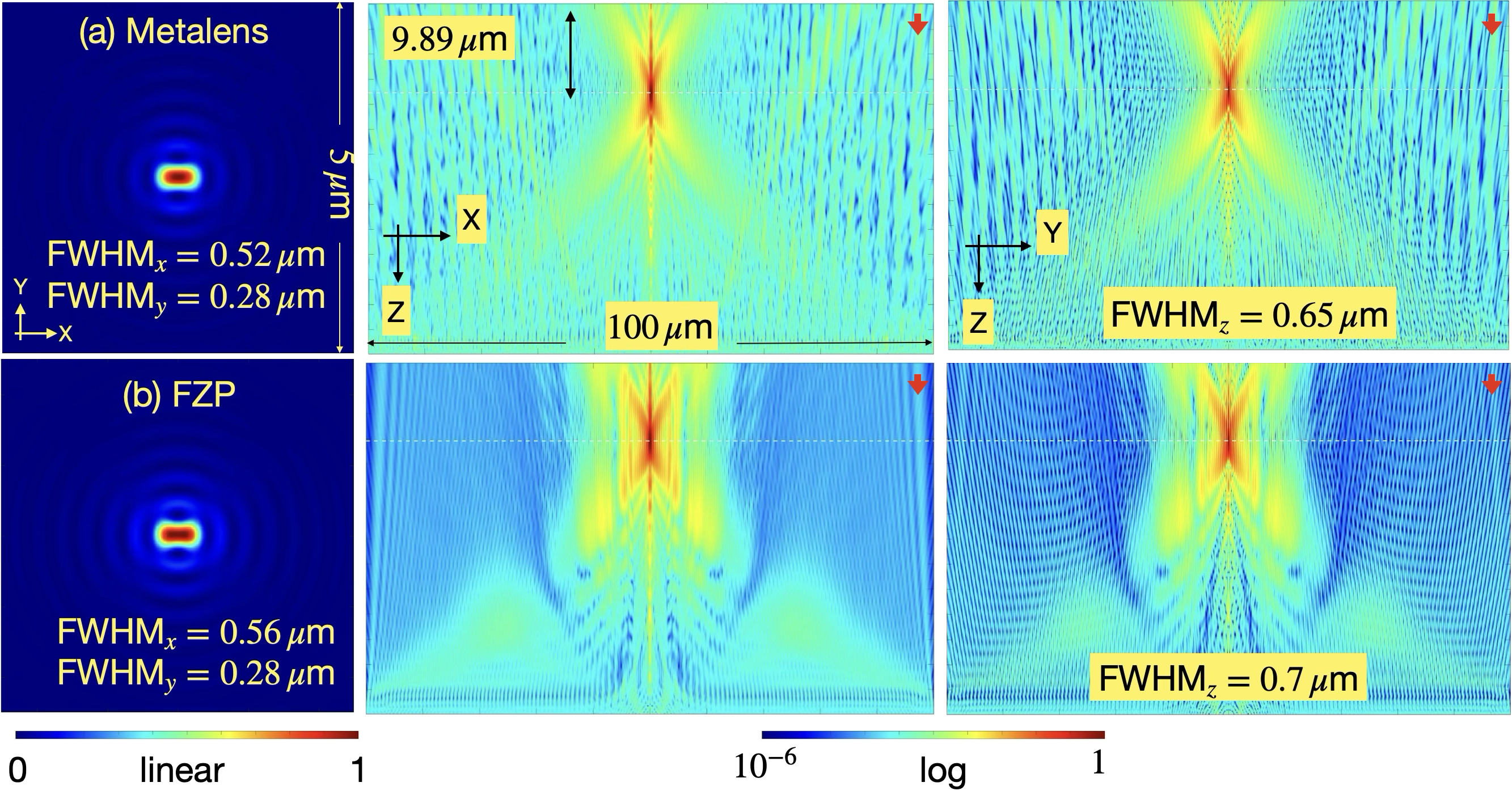}
\caption{\label{fig:Fig1} Focusing comparison between (a) a conventional metalens and (b) a Fresnel Zone Plate (FZP), shown through intensity distributions in the XY plane (linear scale) and in the XZ and YZ planes (log scale). The lenses are positioned at the top of the XZ-, YZ-plane images, illuminated by TE-polarized light (electric field aligned along the X-axis) propagating in the +Z-direction. The transverse focal spot is elliptical with the long axis oriented along the polarization direction of the incident light. Red arrows indicate direction of incident light.}
\end{figure}

\begin{figure}[htbp]
\centering\includegraphics[width=10cm]{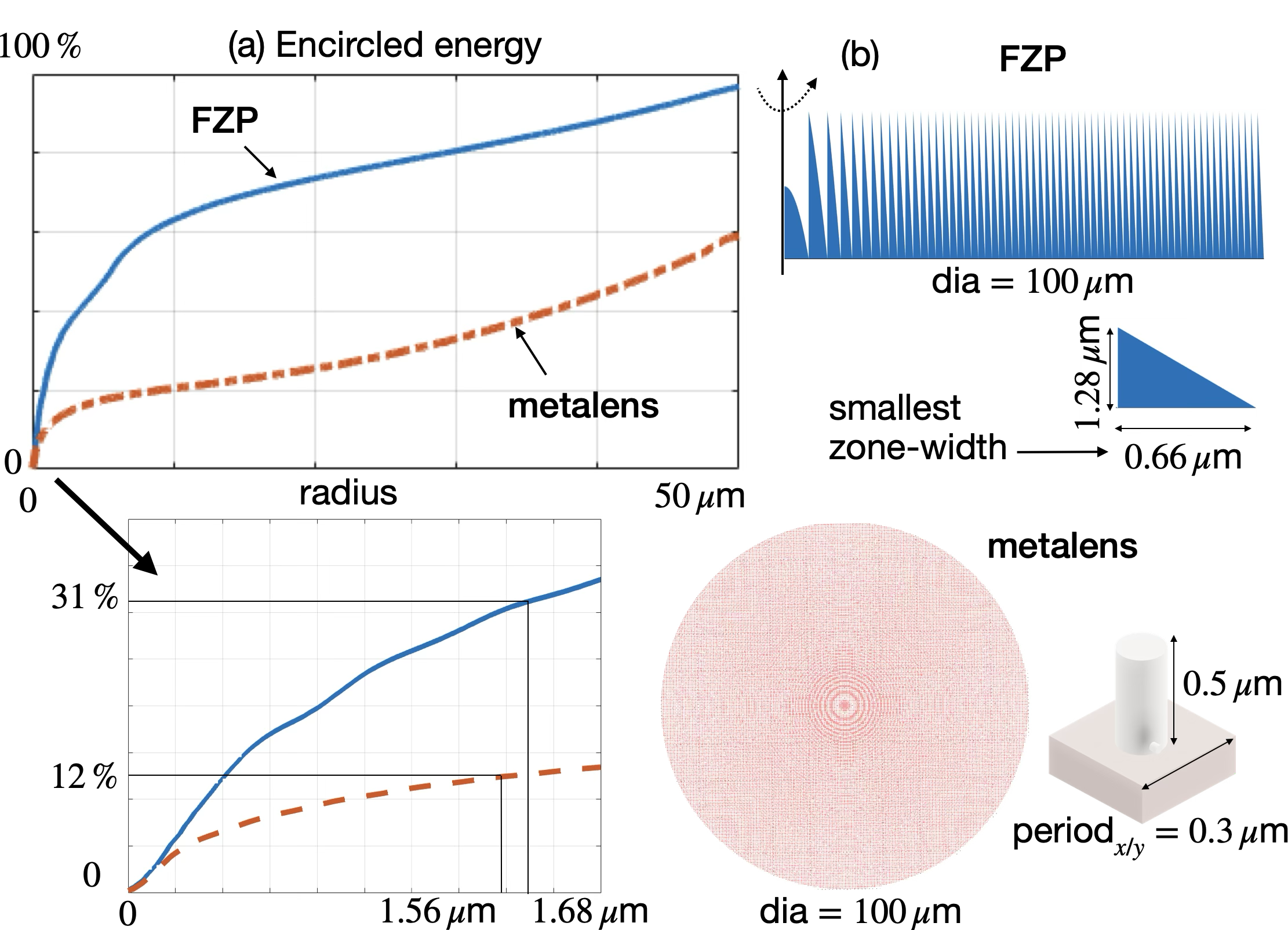}
\caption{\label{fig:Fig2} (a) Encircled energy plots highlight the higher focusing efficiency of the FZP compared to the metalens. The bottom inset shows a magnified view, with $3\times \text{FWHM}_x$ radii clearly marked for both the metalens and the FZP. (b) Geometries of the radially symmetric FZP and the pillar-based metalens.}
\end{figure}

\begin{figure}[htbp]
\centering\includegraphics[width=9cm]{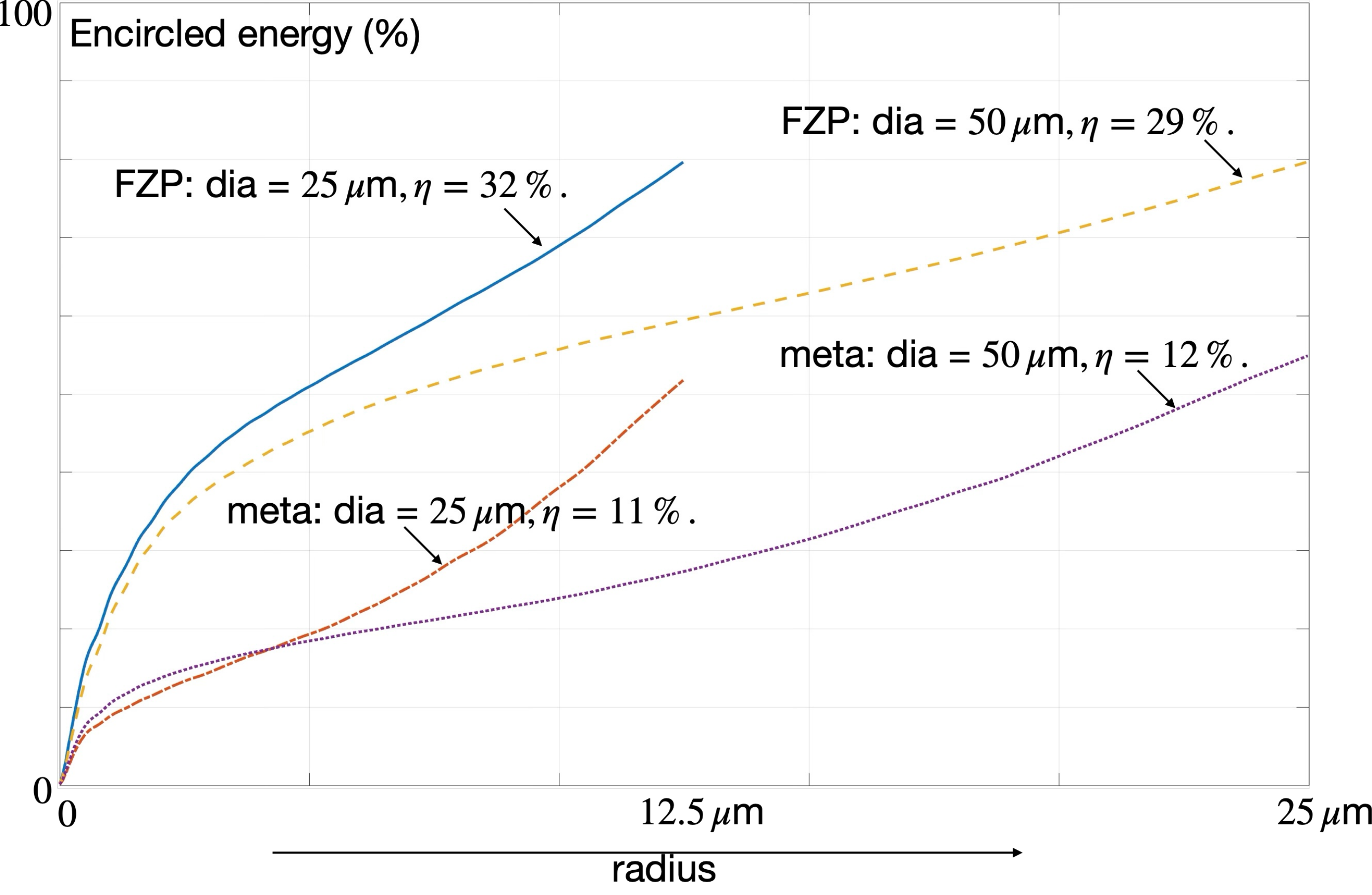}
\caption{\label{fig:EE_25_50}Encircled energy plots for metalens and FZP with diameters of $25 \thinspace \mu$m and $50 \thinspace \mu$m, demonstrating that the FZP consistently outperforms the metalens in focusing efficiency, regardless of size. Efficiency is measured at a radius of $3\thinspace \times\thinspace$FWHM. All simulation parameters remain unchanged from previous figures.}
\end{figure}

In this study, we conduct detailed simulations to compare the focusing efficiency of an FZP and a metalens, both designed with identical specifications at a high numerical aperture (NA = 0.98). Our results demonstrate that the FZP significantly outperforms the metalens, achieving a focusing efficiency of 30\% compared to 12\% (see Figs.~\ref{fig:Fig1} and \ref{fig:Fig2}). This performance advantage persists across different lens diameters of $25\,\mu$m, $50\,\mu$m, and $100\,\mu$m, all maintaining the same NA (see Fig.~\ref{fig:EE_25_50} and Supplement 1). These findings directly challenge prevailing claims in the literature that metalenses offer superior focusing efficiency. Theoretical limits further suggest that conventional metalenses with NA $> 0.9$ inherently underperform.\cite{sang2022toward} 

For clarity, we do not consider optimization techniques such as inverse design.\cite{menon2023perspectives} Additionally, reflection losses are excluded from our analysis, as they can be effectively mitigated with appropriate anti-reflection coatings, particularly for operation at a single wavelength.

At an even higher NA = 0.99, prior studies suggest that topological optimization, leading to so-called \emph{freeform} metalens design yields the highest focusing efficiency compared to conventional unit-cell-based metalenses.\cite{sang2022toward} For instance, theoretical limits predict a focusing efficiency of approximately 30\%, while the best reported freeform metalens achieves only $\sim$22\% (only in simulations, see Table S1 in \cite{sang2022toward}). Here, we emphasize that such freeform metalenses are extremely challenging to fabricate. In contrast, the conventional metalens exhibits a focusing efficiency of merely $\sim$8.25\% (with a transmission efficiency of $\sim$55\% and a diffraction efficiency of $\sim$15\%). Notably, the definition of focusing efficiency appears to differ in this case, where it is reported as the fraction of incident light focused within a spot of radius $1.5\times\text{FWHM}$, rather than $3\times\text{FWHM}$ (see Fig.~2f and Eq.~(3) in \cite{sang2022toward}). Transmission efficiency is the ratio of the transmitted light power to the incident light power through the lens. 

A critical limitation of the metalens design in \cite{sang2022toward} is its reliance on intricate geometries with feature sizes as small as 5 nm and a height of 600 nm in TiO$_2$, yielding an impractical aspect ratio of 30:1. Furthermore, the authors did not compare the performance of an FZP with identical specifications. In this work, we address this gap by providing a direct performance comparison under identical design constraints.

First, we simulated two flat lenses—a FZP and a conventional metalens—each with a diameter of $100\thinspace\mu$m and a focal length of $9.89 \thinspace\mu$m, corresponding to NA of 0.98, optimized for focusing $\lambda=647\thinspace$nm. Both lenses were modeled using the same dielectric material, underlying substrate and identical simulation parameters in RSoft (Synopsys). Details including our validation results are accessible on the project's GitHub page,\cite{fzp_metalens_github} and in Supplement 1. The results are summarized in Fig. \ref{fig:Fig1}. Simulations were performed in 3D, using a normally incident TE-polarized (electric field oriented parallel to the X-axis) plane wave propagating along the Z-axis. In each case, the is lens positioned at the top of the computational domain. The resulting intensity distributions for the XY (linear-scale), XZ, and YZ (log-scale) planes are shown in Figs. \ref{fig:Fig1}(a) and \ref{fig:Fig1}(b) for the metalens and FZP, respectively. As expected,\cite{menon2006experimental, chao2005immersion} the transverse focal spot has an elliptical shape, with its long axis aligned with the polarization direction of the incident light. Based on these simulations, the encircled energy—defined as the fraction of transmitted light focused within a specified radius centered on the focus—was calculated (see Fig. \ref{fig:Fig2}a). The results clearly show that the FZP achieves significantly higher efficiency compared to the metalens. The geometries of both lenses are illustrated in Fig. \ref{fig:Fig2}b. The focusing efficiencies were calculated from the encircled energy plots using the standard definition: the ratio of the transmitted light focused within a radius of $3 \thinspace\times$ the full-width at half-maximum (FWHM$_\text{x}$) in the same polarization direction as the incident light, consistent with the definition in Ref. \cite{sang2022toward}. At this radius, the FZP achieved a focusing efficiency of \textbf{30\%}, significantly outperforming the metalens, which achieved only \textbf{12\%}. At a smaller radius of $1.5\thinspace\times$ FWHM$_\text{x}$, the efficiencies were 20\% for the FZP and 9\% for the metalens.  

This comparison was repeated for lenses with diameters of $25\thinspace\mu$m and $50\thinspace\mu$m, each with NA of 0.98. Figure \ref{fig:EE_25_50} presents encircled energy plots for both the metalens and the FZP with diameters of $25 \thinspace \mu$m and $50 \thinspace \mu$m, demonstrating that the FZP consistently achieves higher focusing efficiency across different sizes (see additional results in Supplement 1). This result confirms the generality of our conclusions, as the performance advantage of the FZP is independent of diameter. To understand the underlying cause of this difference, we conducted a detailed investigation and invoke the shadow effect as described next. 

\section{The Shadow effect} 
\begin{figure}[htbp]
\centering\includegraphics[width=9cm]{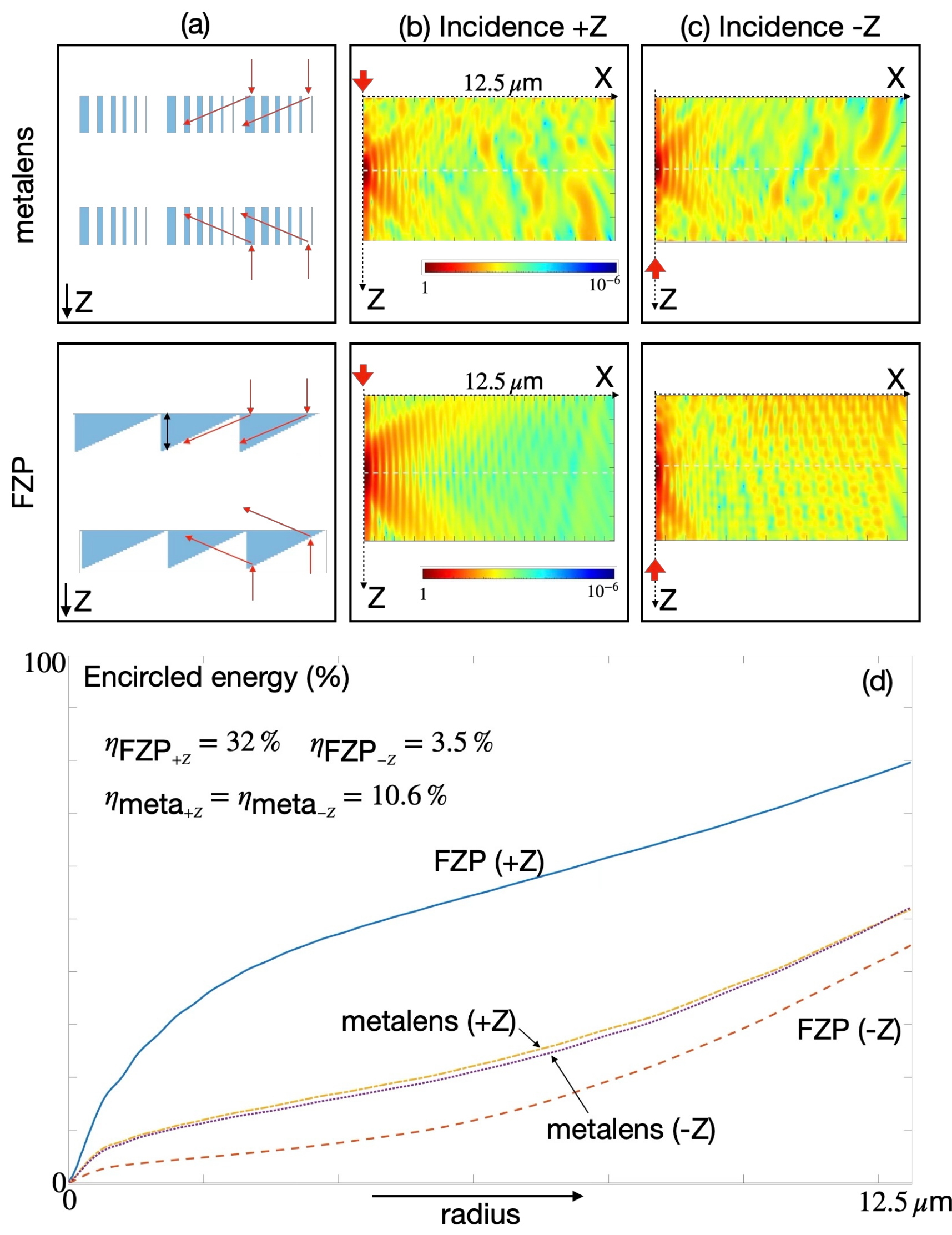}
\caption{\label{fig:shadow} Comparison of shadow effects between a metalens (top row) and a FZP (bottom row) due to symmetry breaking along the propagation (Z) axis. (a) Diffracted light is partially obscured by neighboring features in both cases, but this shadow effect is more pronounced for the metalens and remains independent of incidence direction. In contrast, the FZP exhibits strong suppression of the shadow effect when light is incident in the +Z direction. (b, c) Simulated intensity distributions in the XZ plane for a metalens and FZP (diameter = $25\thinspace \mu$m, NA $\approx$ 0.98) under +Z and -Z incidence. (d) Encircled energy plots confirm symmetry breaking in the FZP, which achieves higher focusing efficiency for +Z incidence. Simulations assume no substrate, but similar trends are observed with a substrate (see Supplement 1).}
\end{figure}

The primary difference between the two flat lens implementations (FZP vs. metalens) stems from symmetry breaking in the direction of light propagation for the FZP. As shown in Fig. \ref{fig:shadow}a, light incident from the +Z direction interacts differently with the FZP compared to incidence from the -Z direction, whereas the metalens exhibits identical behavior in both directions. Consequently, the metalens is affected by the shadow effect—where diffracted light is partially blocked by neighboring features—in both incidence directions equally. In contrast, the FZP experiences reduced shadowing when illuminated from the +Z direction, as illustrated in Fig. \ref{fig:shadow}a.

To quantify this effect, we simulated the FZP and the metalens without an underlying substrate for both light-incidence directions. The results are summarized in Fig. \ref{fig:shadow}. As intuitively expected, the shadow effect is significant for the metalens in both directions, whereas it is minimized for the FZP when light is incident from the +Z direction. This is evident from the scattered intensity far from the optical axis, which serves as an indicator of light deviating from the desired focusing diffraction order (see Figs. \ref{fig:shadow}b and \ref{fig:shadow}c). 

To further quantify this effect, we analyzed the encircled energy plots shown in Fig. \ref{fig:shadow}d, from which we extracted the focusing efficiencies:  
\textbf{Metalens}: $\eta_{+Z} = \eta_{-Z} = 10.6\%$,  \textbf{FZP}: $\eta_{+Z} = 32\%, \eta_{-Z} = 3.5\%$.  

As expected, the metalens maintains symmetry along the Z-axis, resulting in minimal variation in its focusing efficiency regardless of the light incidence direction. In contrast, the FZP exhibits a significant drop in focusing efficiency when light is incident from the -Z direction due to the pronounced shadow effect. While this outcome aligns with physical intuition, it has not been previously documented.

This finding has important practical implications for the design and application of flat optics. Specifically, the efficiency of the FZP is highly dependent on the direction of light incidence, emphasizing the need for careful choice of FZP orientation to achieve optimal performance. 

\section{Experiments}
\begin{figure}[htbp]
\centering\includegraphics[width=11.4cm]{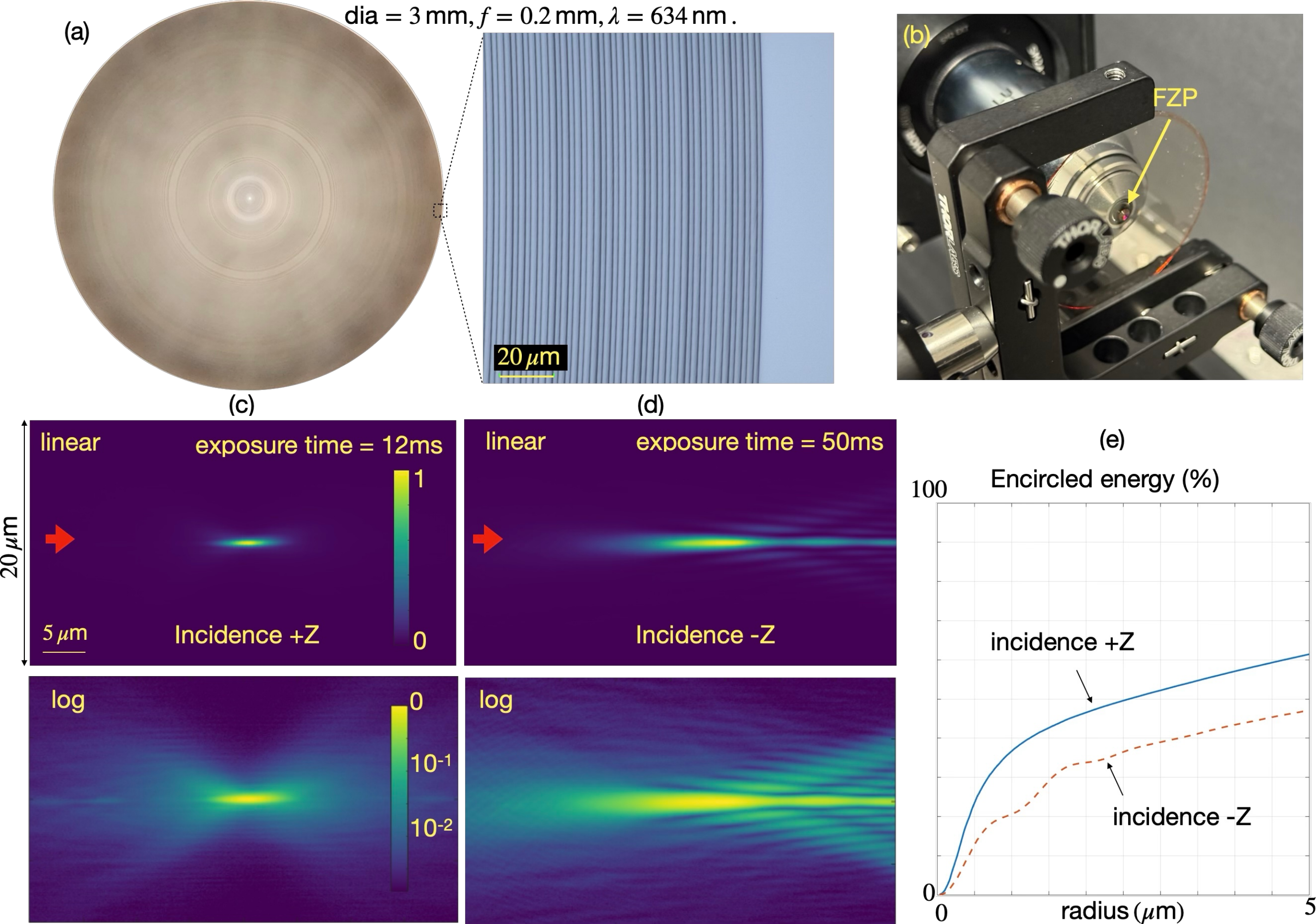}
\caption{\label{fig:expt} Experimental characterization of a high-NA FZP. (a) Optical micrographs of the fabricated FZP (diameter: 3 mm, focal length: 0.2 mm) designed for $\lambda$ = 634 nm. (b) Photograph of the experimental setup, highlighting the high-NA refractive microscope objective (NA = 0.9, $100\times$, working distance = 1 mm) used for point-spread function (PSF) measurements. (c, d) Measured PSFs for illumination in the +Z and -Z directions, respectively, shown in both linear (top) and logarithmic (bottom) intensity scales. (e) Encircled energy plots derived from the recorded PSFs, demonstrating significantly higher focusing efficiency for illumination from the +Z direction.}
\end{figure}

We fabricated a FZP with a 3 mm diameter and a designed focal length of 0.2 mm, yielding a designed (geometric) NA of 0.99 operating at $\lambda=$634 nm, using grayscale optical lithography (GOL). In this process, a 2.5 $\mu$m-thick layer of Shipley 1813 photoresist was spin-coated onto a 0.55 mm-thick sodalime-glass wafer (University Wafers) and patterned using the high-resolution write head of a DWL 66+ laser lithography system (Heidelberg Instruments GmbH). By varying the exposure dose, we precisely controlled the resist height to achieve a blazed FZP profile.\cite{hayward2023multilevel, hayward2024imaging, lin2023inverse, meem2020large} The design was discretized into rings of 0.5 $\mu$m width, with heights ranging from 0 to 1.2 $\mu$m, corresponding to $\lambda$/(n-1). Optical micrographs of the fabricated device are shown in Figs. \ref{fig:expt}(a) and \ref{fig:expt}(b). We note that we were unable to simulate the performance of this FZP due to its large diameter. 

The focusing performance of the fabricated FZP was characterized by illuminating the device with a collimated plane wave from a supercontinuum source (NKT Photonics). A tunable filter was used to select a center wavelength of 634 nm with a 5 nm bandwidth. The experimental setup follows our previously reported configuration.\cite{meem2020large} The point-spread function (PSF) was recorded using a high-NA (0.9, $100\times$, working distance = 1 mm) refractive microscope objective(Nikon LU Plan Fluor), coupled with a tube lens (Thorlabs ITL200, f = 200 mm). Measurements were performed for both FZP orientations under identical experimental conditions, with results summarized in Figs. \ref{fig:expt}(c) and \ref{fig:expt}(d) for illumination in the +Z and -Z directions, respectively. 

As predicted, the FZP exhibits pronounced asymmetry in focusing performance. Under +Z illumination, the PSF is more tightly focused (FWHM = 479 nm vs. 783 nm), background noise is significantly reduced (see log-scale plots in Fig. \ref{fig:expt}c), and the required sensor exposure time is markedly lower (12 ms vs. 50 ms). The FWHM is primarily constrained by the 500 nm ring width, dictated by fabrication limits, effectively restricting the device’s operational NA to 0.66 (+Z) and 0.4 (-Z). Simulations of a smaller FZP further confirm this constraint (see Supplement 1).

Encircled energy analysis, computed from the recorded PSFs, highlights the significantly higher focusing efficiency in the +Z direction (Fig. \ref{fig:expt}d). At a radius of $3\thinspace\times\thinspace$FWHM (+Z direction), the focusing efficiency is approximately 60\% higher. While absolute efficiency measurements are limited by the microscope objective’s field-of-view,\cite{meem2020large} the relative comparison remains rigorous. These experimental results align well with our prior simulations of smaller FZPs, further underscoring the critical role of illumination direction in optimizing the performance of high-NA diffractive optics.  

\section{Discussion}
To the best of our knowledge, this work provides the first demonstration that a FZP can outperform a metalens in focusing efficiency at high NA when its orientation is optimized. We attribute this superior performance to the inherent asymmetry of the FZP along the optical axis, which reduces the shadow effect in the +Z direction—a phenomenon absent in metalenses, where shadowing persists in both directions. Our experiments further confirm this asymmetry in a high-NA FZP. These findings challenge the prevailing assumption that metalenses inherently achieve superior focusing efficiency at high NA, demonstrating instead that such claims are not only unsubstantiated, but also incorrect. 

FZPs are often criticized for their 3D geometry, but Swanson showed in the 1980s that they can be fabricated using standard semiconductor lithography.\cite{swanson1989binary} Advances in nanoimprint lithography now enable cost-effective, high-volume production of 3D diffractive optics in low-refractive-index dielectrics.\cite{majumder2022broadband} In contrast, the metalens in ref. \cite{sang2022toward} required impractical 5 nm features with a 600 nm height, resulting in a 30:1 aspect ratio.

\begin{figure}[htbp]
\centering\includegraphics[width=11.4cm]{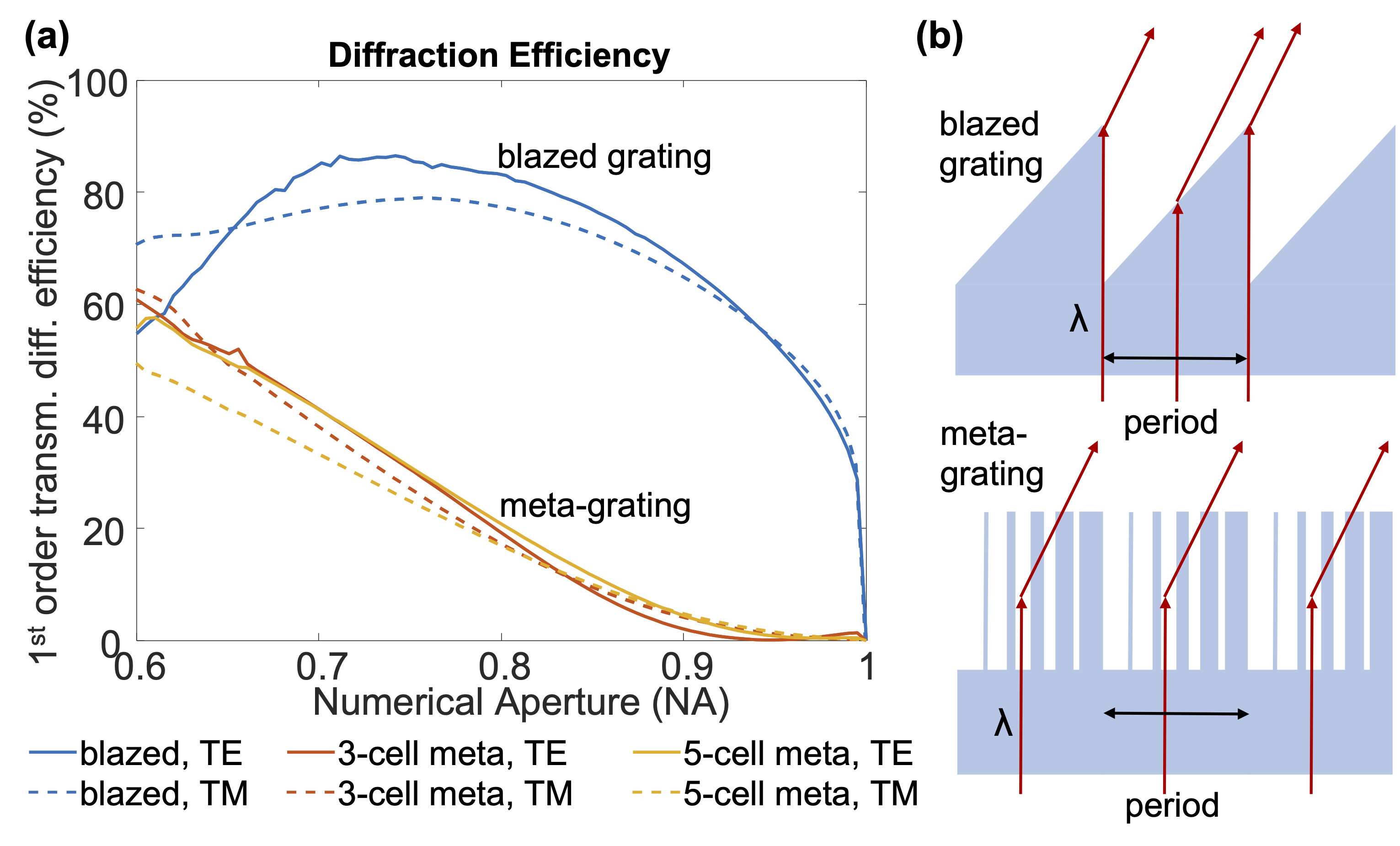}
\caption{\label{fig:Gratings} First-order diffraction efficiencies (a) for blazed and meta-gratings in TE and TM polarizations, with their respective geometries illustrated on the right (b). Notably, the shadow effect is minimized for blazed gratings when light is incident from the substrate side (see section 6, Supplement 1).\cite{fzp_metalens_github}}
\end{figure}

Claims of higher diffraction efficiency have also been made for metasurfaces, \cite{zhou2024large} but to our knowledge, no direct comparisons with conventional diffractive optics or computer-generated holograms have been reported for high-NAs. To explore this, we simulated the diffraction efficiencies of a blazed (prismatic) grating and a conventional meta-grating \cite{wang2024review}, finding that the blazed grating significantly outperforms the meta-grating as summarized in Fig. \ref{fig:Gratings}. Details of our simulations and validation steps are provided in the project's github page\cite{fzp_metalens_github} and in Supplement 1. This result contrasts with most recent publications. While topology optimization could improve both designs, our findings show no inherent advantage of a meta-grating over a blazed grating. Since any linear optical device can be expressed as a linear combination of gratings,\cite{miller2012all} we emphasize that our analysis here is completely general.

Our findings emphasize the importance of rigorous evaluation of claims in the \emph{meta-optics} field. While meta-optics offers unique advantages, such as polarization control,\cite{shen2014ultra} claims of superior performance, especially in focusing efficiency, must be substantiated through direct comparisons with conventional diffractive optics. Our work highlights the limitations of metalenses at high numerical apertures, urging a balanced perspective on the capabilities and trade-offs of meta-optical devices.

\begin{backmatter}
\bmsection{Funding}
Funding from Office of Naval Research (N00014-22-1-2014) is gratefully acknowledged. 

\bmsection{Acknowledgment}
We thank Ke Liu (Synopsis) for assistance with RSoft and Prof. S. Blair and Dr. S. N. Qadri for use of the high-NA objectives for PSF measurements. 

\bmsection{Disclosures}
\noindent HS: Lumarray (I,P). RM: Oblate Optics (I,E,P).

\bmsection{Data Availability Statement}
Supporting data and code underlying the results presented in this paper are available on Github.\cite{fzp_metalens_github}.

\bmsection{Supplemental document}
See Supplement 1 for supporting content.
\end{backmatter}


\bibliography{Main_text}

\end{document}